\documentclass[10pt,journal,twocolumn]{IEEEtran}
\IEEEoverridecommandlockouts
\usepackage{epsfig,cite,bm,algorithm,algorithmic,epstopdf,amsmath,subfigure,multirow,amssymb,amsfonts,amsthm,xcolor,caption,graphicx,textcomp,xcolor, subfigure, makecell, bbding, soul}
\usepackage[inline, shortlabels]{enumitem}
\usepackage[flushleft]{threeparttable}
\soulregister{\cite}7
\soulregister{\ref}7
\setlist{itemjoin ={,\enspace},itemjoin* = { and\enspace}}

\usepackage[T1]{fontenc}
\usepackage{stfloats}
\allowdisplaybreaks[4]

\allowdisplaybreaks[4]
\usepackage{booktabs}

\begin{document}

\title{Hybrid Beamfocusing Design for RSMA-Enhanced Near-Field Secure Communications}
	
\author{Jiasi Zhou, Huiyun Xia, Chuan Wu, and Chintha Tellambura,~\IEEEmembership{Fellow,~IEEE}
\thanks{Jiasi Zhou and Chuan Wu are with the School of Medical Information and Engineering, Xuzhou Medical University, Xuzhou, 221004, China, (email: jiasi\_zhou@xzhmu.edu.cn and 100002018005@xzhmu.edu.cn). (\emph{Corresponding author: Chuan Wu}).}
\thanks{Huiyun Xia is with the Jiangsu Key Laboratory of Wireless Communications, Nanjing University of Posts and Telecommunications, Nanjing 210003, China (email: xiahy2024@njupt.edu.cn).}
\thanks{ Chintha Tellambura is with the Department of Electrical and Computer Engineering, University of Alberta, Edmonton, AB, T6G 2R3, Canada (email: ct4@ualberta.ca).} 
\thanks{This work was supported by the Talented Scientific Research Foundation of Xuzhou Medical University (D2022027).}}
\maketitle

\begin{abstract}
Near-field spherical wavefronts enable spotlight-like beam focusing to mitigate unintended energy leakage, creating new opportunities for physical-layer security (PLS). However, under hybrid analog–digital (HAD) antenna architectures, beamfocusing alone may not provide foolproof privacy protection due to reduced focusing precision. To address this issue, this paper proposes a rate-splitting multiple access (RSMA)-enhanced secure transmit scheme for near-field communications with fully-connected or sub-connected HAD architectures. In the proposed scheme, the common stream is designed for dual purposes, delivering the desired message for legitimate users while acting as artificial noise to disrupt eavesdropping. The primary objective is to maximize the minimum secrecy rate by jointly optimizing the analog beamfocuser, digital beamfocuser, and common secrecy rate allocation. To solve the formulated non-convex problem, we develop a penalty-based alternating optimization algorithm. Specifically, the variables are partitioned into three blocks, where one block is solved via a surrogate optimization method, while the others are updated in closed form. Simulation results reveal that our transmit scheme: (1) approaches fully digital beamfoucsing with substantially fewer radio frequency chains, (2) outperforms conventional beamfocusing-only and far-field security schemes, and (3) preserves secrecy without significantly compromising communication rates.
\end{abstract} 


\section{Introduction} 
The broadcast nature of wireless transmission media makes communicated data susceptible to eavesdropping, raising critical security and confidentiality concerns. Physical layer security (PLS) addresses these threats by harnessing the intrinsic randomness of wireless channels\cite{10478949}. Unlike traditional cryptographic approaches, PLS establishes secure communication without complex secret key management, providing a lightweight yet effective security paradigm. This distinctive advantage has stimulated extensive research into PLS-based solutions for safeguarding far-field communications (FFC)\cite{10478949,10285055,9792580,10311400}, where the electromagnetic wavefronts are approximated as planar. 

However, to support autonomous driving, extended reality, and smart infrastructure, six-generation (6G) wireless networks are transitioning toward extremely large-scale antenna arrays (ELAAs) and high-frequency spectra\cite{10558818}. These shifts fundamentally alter electromagnetic characteristics, rendering the planar wave assumption invalid. Instead, near-field spherical wavefronts become dominant, introducing distance-dependent channel variations alongside angular information. This dual-dimensional characteristic enables the joint exploitation of both distance and direction information. Consequently, the enhanced spatial resolution concentrates radiated energy on specific spatial coordinates, surpassing the limitations of conventional angular beamforming\cite{10663521}.

This spotlight-like beamfocusing effect inherently limits unintended energy leakage, strengthening security even when eavesdroppers share the same angular direction as legitimate users\cite{10436390}. Such protection is theoretically unachievable in FFC systems that rely solely on angular discrimination. As such, spherical wavefronts unlock new opportunities for advancing PLS paradigms. However, current PLS frameworks are predominantly built on planar wave assumptions, creating a critical mismatch with real-world wireless environments\cite{10478949,10285055,9792580,10311400}.

Realizing high-precision beamfocusing ideally requires fully digital antenna architectures, where each antenna element is connected to an independent radio frequency (RF) chain \cite{10123941}. In near-field communication (NFC) scenarios employing ELAAs, such configurations prove impractical due to exorbitant hardware expenses and power consumption. As a more feasible alternative, hybrid analog-digital (HAD) antenna architectures are widely adopted to reduce implementation complexity\cite{zhou2025sub}. This architectural compromise diminishes beam focusing accuracy, causing energy dispersion into undesired regions\cite{zhou2025sub}. Consequently, beamfocusing alone may not provide a foolproof barrier against malicious attacks, necessitating additional countermeasures such as artificial noise\cite{zhao2025near,10480457}. These protective measures, while effective, degrade the channel capacity available to legitimate users and consume valuable transmit power\cite{10415047}.

To address these challenges, rate-splitting multiple access (RSMA) has emerged as a versatile transmission strategy. In RSMA, the base station (BS) splits each user’s message into a common part and a private part. The common parts are jointly encoded into a single common stream, while private parts are encoded into dedicated streams. By tuning the message-splitting ratio, the BS enables users to decode part of the multi-user interference while treating the remainder as noise, thereby achieving flexible and robust interference management. This tunable framework subsumes both space-division multiple access (SDMA) and non-orthogonal multiple access (NOMA) as special cases \cite{mao2018rate,10273395}. Crucially, the common stream can be exploited for \textbf{dual functionality}: conveying data to legitimate users while simultaneously serving as an\textbf{ intentional jamming} signal against eavesdroppers, thus reducing both power consumption and information leakage \cite{10945425,zhang2025fluid}. However, despite this potential, the dual role of RSMA in ensuring physical-layer security remains largely unexplored in the context of secure NFC. 

\subsection{Related Works}
\subsubsection{Beamfocusing-based PLS for NFC}
The potential of exploiting beamfocusing to mitigate privacy leakage has been preliminarily explored in \cite{10436390,10971913,liu2025physical,zhang2024near,10902048,11018844,zhao2025near,10480457}. A central finding in \cite{10436390} indicates that secrecy performance depends primarily on the distance disparity between the eavesdropper and the legitimate user, rather than their angular separation. Extending this result, \cite{10971913} derives closed-form secrecy capacity expressions under three distinct near-field channel models, showing that beamfocusing significantly enlarges the secure transmission region, especially when eavesdroppers are angularly aligned with legitimate users. Building on this foundation, \cite{liu2025physical} investigates a mixed far-field/near-field secure communication scenario, while \cite{zhang2024near} examines wideband secure NFC via analog beamfocusing. In parallel, \cite{10902048} proposes a far-to-near successive interference cancellation (SIC) decoding scheme for NOMA-enhanced NFC, and \cite{11018844} leverages integrated sensing and communication (ISAC) to strengthen near-field PLS against mobile eavesdroppers. Collectively, these studies demonstrate the effectiveness of near-field beamfocusing for secrecy enhancement. However, when eavesdroppers are located in close proximity to legitimate users, beamfocusing alone becomes insufficient. To overcome this limitation, artificial-noise-aided NFC schemes have been proposed in \cite{zhao2025near,10480457}.

\subsubsection{RSMA-enhanced PLS for FFC} RSMA-based transmit schemes have been developed to defend against internal eavesdropping, where each user may attempt to intercept confidential messages intended for others \cite{10542650}. Its effectiveness has been further validated in more complex scenarios, including the coexistence of both internal and external eavesdroppers\cite{10646391}, and the presence of colluding and non-colluding adversaries\cite{10720715}. A more challenging environment is considered in \cite{salem2024robust}, where the eavesdropper resides within a certain region, but its exact position remains unknown. To counter this spatial uncertainty, artificial noise is injected into the transmitted signals, albeit at the cost of increased transmit power. In pursuit of energy-efficient PLS, the dual use of the RSMA common stream is explored in\cite{10817512}. Simulations demonstrate that RSMA achieves considerable secrecy gain over NOMA and SDMA. The RSMA-based security solutions have been successfully extended to ISAC systems, such as those empowered by reconfigurable intelligent surface (RIS)\cite{10945425} and fluid antenna arrays\cite{zhang2025fluid}. However, all these contributions are limited to far-field PLS scenarios with fully digital antenna architectures\cite{10542650,salem2024robust,10817512,10945425,zhang2025fluid}.

\subsubsection{RSMA-enabled NFC without PLS considerations} To better manage multi-user interference, increasing attention has been directed towards RSMA-enabled NFC. The authors in \cite{11071287} investigate the performance of RSMA in NFC with imperfect channel state information (CSI) and SIC. Under similar assumptions, reference\cite{zhou2025sub} evaluates the beamfocusing capability in reducing energy leakage to surrounding users. The results indicate that, even with perfect CSI, beamfocusing alone cannot fully suppress leakage, implying potential eavesdropping risks. Interestingly, the leaked energy can be repurposed to support additional users\cite{10798456} or to assist target sensing\cite{10906379}. To reduce hardware cost, a HAD beamfocusing architecture is adopted in RSMA-enabled mixed near- and far-field communications\cite{10414053}. Building on similar frameworks, RSMA-based transmit schemes have also been developed for near-field ISAC, with sensing accuracy evaluated using detection rate \cite{zhou2024hybrid} and Cram\'{e}r–Rao bound (CRB) \cite{zhou2025crb}.

\subsection{Main Contributions}
Against the above background, this paper exploits the dual use of the RSMA common stream for securing NFCs with fully-connected or sub-connected hybrid antenna architectures. The main contributions are summarized as follows:
\begin{itemize}
\item \textbf{Novel Secure Transmit Scheme:} We propose an RSMA-enhanced PLS transmit framework for NFC, incorporating  HAD beamfocusing. The RSMA common stream delivers the desired message for legitimate users and acts as artificial noise to hinder eavesdropping, while the HAD beamfocusing design reduces RF chain requirements. The minimum secrecy rate maximization problem is formulated, which involves the joint optimization of the analog beamfocuser, the digital beamfocuser, and common secrecy rate allocation.

\item \textbf{Algorithm Design:} We develop a penalty-based alternating optimization algorithm. Specifically, after introducing auxiliary variables, the optimization variables are divided into three blocks and optimized in an alternating manner.
\begin{enumerate}
\item \emph{Auxiliary variables and common secrecy rate optimization}: This subproblem is addressed by adopting a surrogate optimization method, where tractable surrogates are constructed to approximate complex legitimate and eavesdropping rates. An iterative algorithm is then developed to solve the reformulated subproblem.
\item \emph{Digital beamfocusing optimization:} The optimal digital beamfocusing is optimized in closed form.
\item \emph{Analog beamfocusing optimization:} For the fully-connected architecture, an element-wise optimization strategy is employed, with each element derived in closed form. For the sub-connected architecture, the optimal analog beamfocusing is obtained in closed form by exploiting its block-diagonal structure. 
\end{enumerate}

\item \textbf{Numerical Insights:} Extensive simulations highlight three key advantages of our proposed transmit scheme over four benchmarks: (1) achieving secrecy performance comparable to fully digital beamfocusing with fewer RF chains, (2) providing substantial gains over beamfocusing-only and far-field schemes, and (3) ensuring secure transmission without significantly sacrificing communication rates. 
\end{itemize}

\emph{Organization:}  The remainder of this paper is organized as follows.  Section \ref{Section II} presents the signal model and formulates the minimum secrecy rate maximization problem. Section \ref{Section III} proposes an efficient iterative algorithm and discusses several crucial properties. Section \ref{Section IV} provides the simulation results. Finally, section \ref{Section V} concludes this paper.

\emph{Notations:} Boldface upper-case letters, boldface lower-case letters, and calligraphy letters denote matrices, vectors, and sets. The complex matrix space of $N\times K$ dimensions is denoted by $\mathbb{C}^{N\times K}$. The superscripts ${(\bullet)}^T$ and ${(\bullet)}^H$ represent the transpose and Hermitian transpose, respectively. $\text{Re}\left( \bullet\right)$, $\text{Tr}\left( \bullet\right)$, and $\mathbb{E}\left[\bullet\right]$ 
 denote the real part, trace, and statistical expectation. $\text{diag}\left( \bullet\right)$  and $\text{blkdiag}\left( \bullet\right)$ denote diagonal and block diagonal operations, respectively. The Frobenius norm of matrix $\mathbf{X}$ is denoted by $||\mathbf{X}||_F$. For matrix $\mathbf{X}$, $\mathbf{X}\left(i:j,:\right)$  represent a sub-matrix composed of the rows from the $i$-th to $j$-th. For vector $\mathbf{x}_i$, $\mathbf{x}_{i,j}$  represent its $j$-th element. Variable  $x\sim\mathcal{CN}(\mu, \sigma^2)$ is a  circularly symmetric complex Gaussian (CSCG) with mean $\mu$ and variance $\sigma^2$. 

\section{System Model and Problem Formulation}\label{Section II}
\begin{figure}[tbp]
\centering
\includegraphics[scale=0.5]{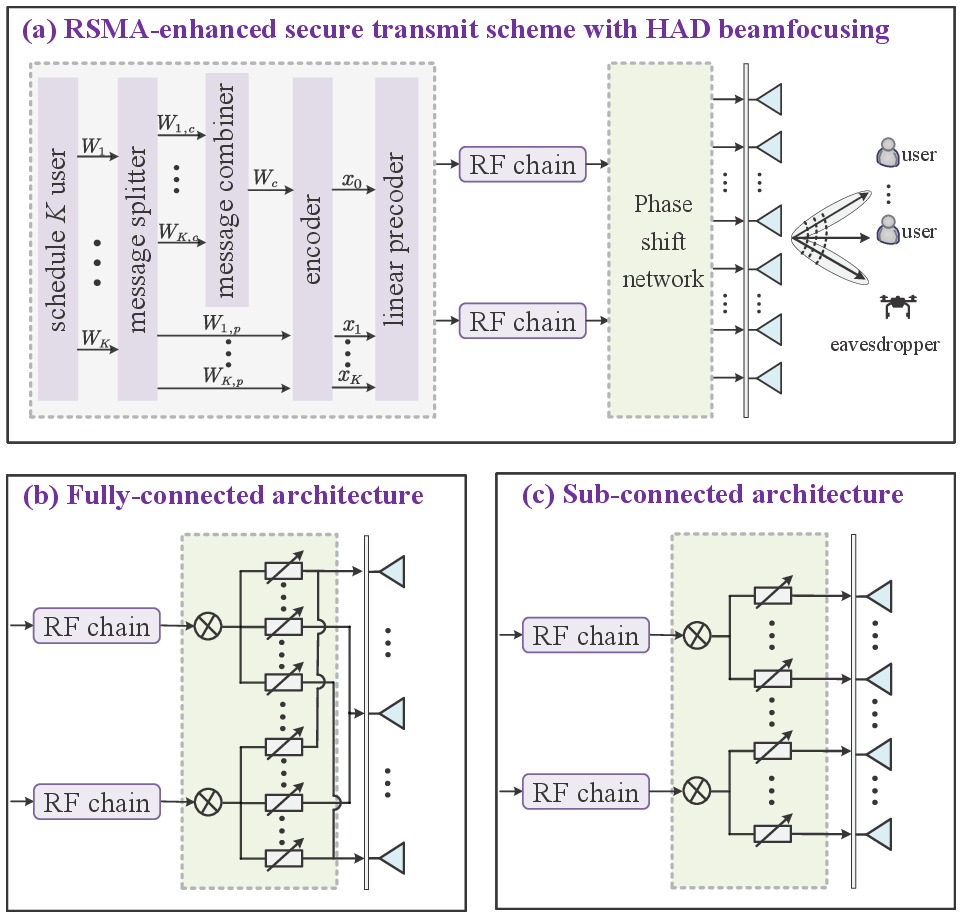}
\caption{The considered RSMA-enhanced secure NFC.}
\label{system}
\end{figure}
We consider an RSMA-enhanced secure NFC system, which comprises a BS equipped with $N$ antennas and $L$ RF chains, $K$ single-antenna legitimate users, and a single-antenna eavesdropper. The set of transmit antennas, RF chains, and users are denoted by $\mathcal{N}=\left\{1,\dots,N\right\}$, $\mathcal{L}=\left\{1,\dots,L\right\}$, and $\mathcal{K}=\left\{1,\dots,K\right\}$, respectively. The BS adopts a uniform linear array (ULA) with an inter-element spacing of $d$. All legitimate users and the eavesdropper are located within the Rayleigh distance $d_{\text{r}}=\frac{2D^2}{\lambda}$, where $D=\left(N-1\right)d$ and $\lambda$ are the antenna aperture and signal wavelength, respectively.

NFC typically operates in ELAA regimes, where RF chain deployment presents significant hardware challenges.  To alleviate hardware overhead, this paper considers two HAD antenna architectures utilizing phase shifters, described as follows:
\begin{itemize}
\item \emph{Fully-connected HAD architecture}: As depicted in Fig.~\ref{system}(b), each RF chain is connected to all transmit antenna elements. Consequently, each entry in $\mathbf{F}\in\mathbb{C}^{N\times L}$ should meet the unit-modulus constraint, i.e.,
\begin{equation}
\mathcal{F}_{1}=\left\{\mathbf{F}\big|\left|\mathbf{F}_{n,l}\right|=1,~n\in\mathcal{N},~l\in\mathcal{L}\right\}.
\end{equation}

\item \emph{Sub-connected HAD architecture}: As shown in Fig.~\ref{system}(c), each RF chain is connected to a dedicated sub-array composed of $M = N/L$ antenna elements. Consequently, the analog beamfocusing matrix $\mathbf{F}\in\mathbb{C}^{N\times L}$ exhibits a block-diagonal structure, expressed as
\begin{equation}
\mathcal{F}_{2}=\left\{\mathbf{F}\big|\mathbf{F}=\text{blkdiag}\left(\mathbf{f}_1,\dots,\mathbf{f}_{L}\right)\in\mathbb{C}^{N\times L}\right\},
\end{equation}
where $\mathbf{f}_l \in \mathbb{C}^{M \times 1}$ represents the phase-shift vector associated with the $l$-th RF chain. Due to hardware limitation, all non-zero entries in $\mathbf{F}$ must meet the unit-modulus constraint, i.e., $\left|\mathbf{f}_{l,m}\right| = 1$ for all $l \in \mathcal{L}$ and $m \in \mathcal{M} = \{1, \dots, M\}$.
\end{itemize}

\subsection{Near-field channel model}
Without loss of generality, we assume the ULA is aligned along the $y$-axis and centered at the origin. As a result, the coordinate of the $n$-th antenna element is given by $\mathbf{t}_n= (0,y_n)$, where $y_n = \left(n-\frac{N+1}{2}\right)d$. Let $(r_k,\theta_k)$ be the polar coordinates of the $k$-th legitimate user, whose Cartesian coordinate is $\mathbf{u}_k = (r_k\cos\theta_k, r_k\sin\theta_k)$. The propagation distance between the $n$-th antenna and the $k$-th legitimate user is 
\begin{equation}
d_{n,k} = \left\|\mathbf{t}_n-\mathbf{u}_k\right\| = \sqrt{r^2_k + y_n^2 - 2r_k y_n \sin\theta_k}.
\end{equation}
Applying the second-order Taylor expansion to $d_{n,k}$, we have $d_{n,k}\approx r_k-\delta_{n,k}$, where the correction term  $\delta_{n,k}=y_{n}\sin\theta_k-y^2_{n}\cos^2\theta_k/2r_k$. According to the Fresnel approximation, the path loss from all antennas to the $k$-th legitimate user is approximately identical\cite{10579914}. Therefore, the pathloss of all links can be approximated by that of the central link, i.e., $\tilde\beta_{k}=\frac{c}{4\pi f r_k}$, where $f$ is the carrier frequency and $c$ is the speed of light. The channel coefficient between the $n$-th antenna and the $k$-th legitimate user is modeled as
\begin{equation}
h_{n,k}= \tilde\beta_{k} e^{-j\frac{2\pi}{\lambda}d_{n,k}}\approx\beta_{k} e^{-j\frac{2\pi}{\lambda}\left(d_{n,k}-r_k\right)}=\beta_{k} e^{j\frac{2\pi}{\lambda}\delta_{n,k}},
\label{Single_Channel}
\end{equation}
where $\beta_{k}=\tilde\beta_{k}e^{-j\frac{2\pi}{\lambda}r_k}$ is the complex channel gain. Stacking all antenna responses, the near-field channel vector from the BS to the $k$-th legitimate user becomes
\begin{equation}
\mathbf{h}_k = \beta_{k}\left[e^{j\frac{2\pi}{\lambda}\delta_{1,k}},\dots,e^{j\frac{2\pi}{\lambda}\delta_{N,k}}\right]^T=\beta_{k}\mathbf{a}(r_k, \theta_k),
\end{equation}
where $\mathbf{a}(r_k,\theta_k) \in \mathbb{C}^{N\times 1}$ denotes the near-field array response vector, which captures both angle- and distance-dependent phase variations.

Similarly,  the near-field channel vector from the BS to the eavesdropper can be modeled as
\begin{equation}
\mathbf{g}_{\text{e}} = \beta_{\text{e}}\left[e^{j\frac{2\pi}{\lambda}\delta_{1,{\text{e}}}},\dots,e^{j\frac{2\pi}{\lambda}\delta_{N,{\text{e}}}}\right]^T=\beta_{\text{e}}\mathbf{a}(r_{\text{e}}, \theta_{\text{e}}),
\end{equation}
where $(r_{\text{e}}, \theta_{\text{e}})$ is the polar coordinates of the eavesdropper and $\beta_{\text{e}}$ is its complex channel gain.

%In realistic scenarios, multi-path propagation may occur due to the presence of $q$ scatterers. Each additional path is characterized by a distinct angle and propagation distance. Let $r_{k,q}$ and $\tilde r_{k,q}$ denote the distances from the BS to the $q$-th scatterer and from the scatter to the $k$-th user, respectively.  The aggregated channel can thus be modeled as
%\begin{equation}
%\mathbf{h}_k = \beta_{k}\mathbf{a}(r_k, \theta_k) + \sum_{q=1}^{Q} \beta_{k,q} \mathbf{a}(r_{k,q}, \theta_{k,q}),
%\end{equation}
%where $\beta_{k,q}=\tilde\beta_{k,q} e^{-j\frac{2\pi}{\lambda}\left(r_{k,q}+\tilde r_{k,q}\right)}$ is the complex channel gains of the $q$-th Non-LoS (NLoS) path. 

\subsection{RSMA-enhanced signal model and performance metric}
As illustrated in Fig.~\ref{system}(a), the message intended for the $k$-th legitimate user is divided into two components: a common part $W_{k,c}$ and a private part $W_{k,p}$. All common parts $\left\{W_{1,c},\dots,W_{K,c}\right\}$ are aggregated and encoded into a common stream $x_0$, while each private part $W_{k,p}$ is independently encoded into a user-specific private stream $x_k$ for $\forall k$. All streams are mutually independent and normalized to unit power, i.e., $\mathbb{E}|x_{k}|^2=1$ for $\forall k \in\tilde{\mathcal{K}}=\{0,1,\dots, K\}$. The resultant stream vector $\left[x_0,x_1,\dots, x_K\right]$ are then precoded by a hybrid beamfocusing matrix represented as $\mathbf{FW}\in\mathbb{C}^{N\times (K+1)}$, where $\mathbf{F}\in\mathbb{C}^{N\times L}$ and $\mathbf{W}=\left[\mathbf{w}_0,\mathbf{w}_1,\dots, \mathbf{w}_K\right]\in\mathbb{C}^{L\times (K+1)}$ denote analog and digital beamfocusers, respectively. Specifically, $\mathbf{w}_0\in\mathbb{C}^{L\times 1}$ and $\mathbf{w}_k\in\mathbb{C}^{L\times 1}$ in $\mathbf{W}$ are digital beamfocusers for the common stream and the $k$-th private stream, respectively. As such, the transmitted signal at the BS is given by $\mathbf{x}=\mathbf{Fw}_0x_0+\sum^{K}_{k=1}\mathbf{Fw}_kx_k$. 

The received signal at the $k$-th legitimate user and the eavesdropper are respectively given as 
\begin{subequations}
\begin{align}
\tilde y_{k}=&\mathbf{h}^H_k\mathbf{F}\mathbf{w}_{0}x_{0}+ \sum\nolimits^K_{i=1}\mathbf{h}^H_k\mathbf{F}\mathbf{w}_{i}x_{i}+n_{k},\\
\tilde y_{\text{e}}=&\mathbf{g}^H_{\text{e}}\mathbf{F}\mathbf{w}_{0}x_{0}+ \sum\nolimits^K_{i=1}\mathbf{g}^H_{\text{e}}\mathbf{F}\mathbf{w}_{i}x_{i}+n_{\text{e}},
\end{align}
\end{subequations}
where $n_{k}\sim \mathcal{CN}\left(0,\sigma^2_{k}\right)$ and $n_{\text{e}}\sim \mathcal{CN}\left(0,\sigma^2_{\text{e}}\right)$ denote additional white Gaussian noise (AWGN) term. As a result, the received power is respectively expressed as~\cite{7555358}
\begin{subequations}\label{power}
	\begin{align}
	&T_{k,c}=\overbrace{{\left|\mathbf{h}^H_{k}\mathbf{Fw}_0\right|}^2}^{S_{k,c}}+\underbrace{\overbrace{{\left|\mathbf{h}^H_{k}\mathbf{Fw}_{k}\right|}^2}^{S_{{k,p}}}+\overbrace{\sum_{i=1,i\neq k}^{K}\left|\mathbf{h}^H_{k}\mathbf{Fw}_i\right|^2+\sigma^2_k}^{I_{{k,p}}}}_{I_{k,c}=T_{k,p}},
	\label{equ:power_legitimate}\\
	&T_{\text{e},c}=\overbrace{{\left|\mathbf{g}^H_\text{e}\mathbf{Fw}_0\right|}^2}^{S_{\text{e},c}}+\underbrace{\overbrace{{\left|\mathbf{g}^H_\text{e}\mathbf{Fw}_{k}\right|}^2}^{S_{{\text{e},k}}}+\overbrace{\sum_{i=1,i\neq k}^{K}\left|\mathbf{g}^H_\text{e}\mathbf{Fw}_i\right|^2+\sigma^2_\text{e}}^{I_{{\text{e},k}}}}_{I_{\text{e},c}=T_{\text{e},k}}.
	\label{equ:power_eavesdropper}
	\end{align}
\end{subequations}

To retrieve the desired message, the $k$-th legitimate user decodes the common stream by treating all private streams as additional noise. The corresponding signal-to-interference-plus-noise ratio (SINR) is $\gamma_{k,c}=S_{k,c}{I^{-1}_{k,c}}$.  After successfully decoding the common stream, the $k$-th user removes it via SIC and proceeds to decode its desired private stream, yielding an SINR of $\gamma_{{k,p}}=S_{k,p}{I^{-1}_{{k,p}}}$. Accordingly, the achievable rates for the common and private streams at legitimate user $k$ are
\begin{equation}
R_{k,c}=\log\left(1 + \gamma_{{k,c}}\right)~\text{and}~R_{k,p}=\log\left(1 + \gamma_{{k,p}}\right).
\end{equation}
However, to ensure the decodability of the common stream, its actual rate $R_c$ should not exceed the minimum channel capacity among all legitimate users, i.e., $R_c=\min_{\forall k}R_{k,c}$. 

Meanwhile, the eavesdropper attempts to decode the common stream. Its corresponding SINR is $\gamma_{{\text{e},c}}=S_{\text{e},c}{I^{-1}_{\text{e},c}}$, which leads to an eavesdropping rate of $R_{\text{e},c}=\log\left(1 + \gamma_{{\text{e},c}}\right)$. Therefore, the resultant secrecy rate for the common stream is $R^s_{c}=\left[R_c-R_{\text{e},c}\right]^+$, where operator ${[x]}^+=\max{\left(x,0\right)}$. Moreover, since the secrecy rate for the common stream is shared by all legitimate users, we have $\sum^{K}_{k=1}R^s_{k,c}\leq R^s_{c}$, where $R^s_{k,c}$ represents the portion of the common secrecy rate used to transmit $W_{k,c}$. When the condition $R_{\text{e},c}<R_c$ is met, the eavesdropper fails to decode the common stream. In such cases, the undecodable common stream can be regarded as artificial noise, further enhancing the confidentiality of private messages. Consequently, eavesdropping capacity for the $k$-th private stream is  $R_{\text{e},k}=\log\left(1+\gamma_{\text{e},k}\right)$, where $\gamma_{\text{e},k}=S_{\text{e},k}{\left(I_{\text{e},k} + S_{\text{e},c}\right)}^{-1}$. The resultant private secrecy rate is $R^s_{k,p} = {[R_{k,p}-R_{\text{e},k}]}^+$. The total achievable secrecy rate of the $k$-th legitimate user is 
\begin{equation}
R^s_{k} = R^s_{k,c} + R^s_{k,p}. 
\label{ssr}
\end{equation} 

\subsection{Problem formulation}
This paper focuses on a hybrid beamfocusing design and common secrecy rate allocation to maximize the minimum secrecy rate among all legitimate users. The optimization problem is then formulated as
\begin{subequations}\label{linear_p}
	\begin{align}
&\max_{\mathbf{F},\mathbf{W},R^s_{k,c} } \min_{\forall k} R^s_k,\label{ob_a}\\
	\text{s.t.}~
	&||\mathbf{FW}||^2_F\leq P_{\text{th}},\label{ob_b}\\
     &R_{k,c}\geq R_{\text{e},c},~\forall k,\label{ob_c}\\
 &\sum_{k=1}^{K}R^s_{k,c} \leq R^s_{c},\label{ob_d}\\
 &R^s_{k,c} \geq 0,~\forall k,\label{ob_e}\\
  &\mathbf{F}\in\mathcal{F}_x,~\forall x\in\{1,2\}, \label{ob_f} 
	\end{align}
\end{subequations}
where $P_{\text{th}}$ is the maximum transmit power budget. (\ref{ob_b}) limits the transmit power requirement. (\ref{ob_c}) ensures that the common stream can serve as artificial noise. (\ref{ob_d}) and (\ref{ob_e}) enforce the common secrecy rate allocation requirement. (\ref{ob_f}) imposes the unit-modulus condition on the analog beamfocuser $\mathbf{F}$, where $x\in\{1,2\}$ indicates different connection modes of RF chains.

Problem (\ref{linear_p}) is intractable to solve optimally due to three technical challenges. First, the objective function is non-smooth and non-convex, invalidating conventional primal-dual optimization methods due to an unknown duality gap. Second, the analog and digital beamfocusing components are strongly coupled, aggravating the difficulty. Third, the unit-modulus constraint further compounds the optimization difficulties. Consequently, the global optimal solution appears elusive.

\section{Proposed algorithm}\label{Section III}
To address the formulated non-convex problem, this section presents a penalty-based alternating optimization algorithm, where the analog and digital beamfoucsers are obtained in closed form. We then analyze the convergence behavior and computational complexity of the proposed algorithm.

To decouple the analog and digital beamfocusers, we introduce unconstrained fully digital beamfocuser $\mathbf{P}=\left[\mathbf{p}_0,\mathbf{p}_1,\dots, \mathbf{p}_K\right]\in\mathbb{C}^{N\times (K+1)}$ as an auxiliary variable. Ideally, the equality $\mathbf{P}=\mathbf{FW}$ must hold, yielding the following equivalent optimization problem:
\begin{subequations}\label{linear_p2}
	\begin{align}
&\max_{\mathbf{P},\mathbf{F},\mathbf{W},R^s_{k,c} } \min_{\forall k} R^s_k,\label{ob_a2}\\
	\text{s.t.}~
    &\mathbf{P}=\mathbf{FW},\label{ob_b2}\\
	&\mbox{(\ref{ob_b})~--~(\ref{ob_f})}. \label{ob_c2} 
	\end{align}
\end{subequations}
In problem (\ref{linear_p2}), when calculating the legitimate and eavesdropping rates, the received power at the $k$-th legitimate user and the eavesdropper should be redefined based on the fully-digital beamfocuser $\mathbf{P}$, given by
\begin{subequations}\label{power_2}
	\begin{align}
&T_{k,c}=\overbrace{{\left|\mathbf{h}^H_{k}\mathbf{p}_0\right|}^2}^{S_{k,c}}+\underbrace{\overbrace{{\left|\mathbf{h}^H_{k}\mathbf{p}_{k}\right|}^2}^{S_{{k,p}}}+\overbrace{\sum_{i=1,i\neq k}^{K}\left|\mathbf{h}^H_{k}\mathbf{p}_i\right|^2+\sigma^2_k}^{I_{{k,p}}}}_{I_{k,c}=T_{k,p}},
	\label{equ:power_legitimate_2}\\
&T_{\text{e},c}=\overbrace{{\left|\mathbf{g}^H_\text{e}\mathbf{p}_0\right|}^2}^{S_{\text{e},c}}+\underbrace{\overbrace{{\left|\mathbf{g}^H_\text{e}\mathbf{p}_{k}\right|}^2}^{S_{{\text{e},k}}}+\overbrace{\sum_{i=1,i\neq k}^{K}\left|\mathbf{g}^H_\text{e}\mathbf{p}_i\right|^2+\sigma^2_\text{e}}^{I_{{\text{e},k}}}}_{I_{\text{e},c}=T_{\text{e},k}}.
	\label{equ:power_eavesdropper_2}
	\end{align}
\end{subequations}
However, the equality constraint (\ref{ob_b2}) makes the direct optimization of hybrid beamfocusers intractable. To address this challenge, we employ the penalty method, which incorporates the equality constraint into the objective function as a penalty term. Then, removing the minimum operator in the objective function, problem (\ref{linear_p2}) can be recast to 
\begin{subequations}\label{linear_p3}
	\begin{align}
&\max_{\mathbf{P},\mathbf{F},\mathbf{W},R^s_{k,c},R^s,R^s_{k,p} } R^s -\frac{1}{\rho}||\mathbf{P}-\mathbf{FW}||^2_F,\label{ob_a3}\\
	\text{s.t.}~
    &||\mathbf{P}||^2_F\leq P_{\text{th}},\label{ob_b3}\\
    &R^s\leq R^s_{k,c}+R^s_{k,p},~\forall k,\label{ob_c3}\\
    &R^s_{k,p} \leq R_{k,p}-R_{\text{e},k},~\forall k,\label{ob_d3}\\
    &\sum_{k=1}^{K}R^s_{k,c} \leq R_{k,c}-R_{\text{e},c},~\forall k,\label{ob_e3}\\
	&\mbox{ (\ref{ob_c}),~(\ref{ob_e}),~(\ref{ob_f})}, \label{ob_f3} 
	\end{align}
\end{subequations}
where operator ${[\bullet]}^+$ is omitted, as this modification preserves the optimality of the solution. The conclusion can be established by contradiction, following the approach in\cite{8709756}. In problem (\ref{linear_p3}), $\rho>0$ denotes penalty factor. As $\rho\to 0$, the solution approaches feasibility with $\mathbf{P}=\mathbf{FW}$. However, an excessively small initial $\rho$ makes the penalty term dominate the objective function, undermining secrecy rare maximization. To mitigate this, the penalty factor is initialized at a relatively large value to generate a suitable initial point and is subsequently decreased until the equality constraint is sufficiently satisfied. This procedure naturally yields a double-loop framework: the inner loop optimizes variables under fixed $\rho$, while the outer loop updates $\rho$ to enforce the feasibility of the final solution.

Given a penalty factor, solving problem (\ref{linear_p3}) remains intractable due to the coupling between beamfocusing matrices. To address this, the variables are partitioned into three groups and updated in an alternating manner, i.e., $\mathcal{Q}_1=\left\{\mathbf{P},R^s_{k,c},R^s,R^s_{k,p}\right\}$, $\mathbf{W}$, and $\mathbf{F}$. This results in three subproblems per iteration,  with the solution procedure for each presented in the following subsections.

\subsection{Subproblem with respect to $\mathcal{Q}_1$}
With the fixed $\mathbf{F}$ and $\mathbf{W}$, the subproblem for updating $\mathcal{Q}_1$ can be rewritten as
\begin{subequations}\label{linear_p4}
	\begin{align}
&\max_{\mathcal{Q}_1} R^s -\frac{1}{\rho}||\mathbf{P}-\mathbf{FW}||^2_F,\label{ob_a4}\\
	\text{s.t.}~
	&\mbox{ (\ref{ob_c}),~(\ref{ob_e}),~(\ref{ob_b3}),~(\ref{ob_c3}),~(\ref{ob_d3}),~(\ref{ob_e3})}. \label{ob_b4} 
	\end{align}
\end{subequations}
Problem (\ref{linear_p4}) involves the difference of two logarithmic functions, making it difficult to directly optimize the fully digital beamfocuser. To overcome this difficulty, we employ surrogate optimization\cite{lange2000optimization}, which replaces the original objective function with a computationally tractable surrogate. To minimize performance loss, the constructed surrogate must closely approximate the original function. To this end, we develop accurate surrogates for the legitimate rate and eavesdropping rate, as follows.

\subsubsection{Surrogate construction for legitimate rate} To ensure the secrecy rate constraint remains a convex set, we should construct lower-bounded concave quadratic surrogates for the legitimate rates $R_{k,c}$ and $R_{k,p}$.  Motivated by\cite{8976409}, these surrogates are formulated as follows
\begin{subequations}\label{Overall_SCA}
\begin{align}
&f_{k,c}\left(\mathbf{P}\right)= \sum_{i=0}^{K}\mathbf{p}^H_{i}\mathbf{x}_{k,c}\mathbf{p}_{i}+2\text{Re}\left(\mathbf{y}_{k,c}\mathbf{p}_{0}\right)+ z_{k,c},
\label{SCA_1}\\
&f_{k,p}\left(\mathbf{P}\right)= \sum_{i=1}^{K}\mathbf{p}^H_{i}\mathbf{x}_{k,p}\mathbf{p}_{i}+2\text{Re}\left(\mathbf{y}_{k,p}\mathbf{p}_{k}\right) + z_{k,p},
\label{SCA_2}
\end{align}
\end{subequations}
where 
\begin{equation}
\begin{split}
\mathbf{x}_{k,\tau}&=-\frac{1}{\ln 2}\mathbf{h}_k\tilde{\mathbf{u}}_{k,\tau} \left(\tilde v_{k,\tau}\right)^{-1}\tilde{\mathbf{u}}^H_{k,\tau}\mathbf{h}^H_k,\\
\mathbf{y}_{k,\tau}&=\frac{1}{\ln 2}\left(\tilde v_{k,\tau}\right)^{-1}\tilde{\mathbf{u}}^H_{k,\tau}\mathbf{h}^H_k,\\
z_{k,\tau}&=-\frac{1}{\ln 2}\left(\tilde v_{k,\tau}\right)^{-1}\left(\sigma^2_{k}\tilde{\mathbf{u}}^H_{k,\tau}\tilde{\mathbf{u}}_{k,\tau}+1\right)\\&-\log \tilde{v}_{k,\tau}+\frac{1}{\ln 2},
\end{split}
\end{equation}
with $\forall\tau\in\{c,p\}$. The auxiliary variables are defined as
\begin{equation}\label{Auxiliary_1}
\begin{split}
&\tilde{\mathbf{u}}_{k,c}=\left(\tilde T_{k,c}\right)^{-1}\mathbf{h}^H_k\tilde{\mathbf{p}}_{0}~\text{and}~v_{k,c}=1-\tilde{\mathbf{u}}^H_{k,c}\mathbf{h}^H_k\tilde{\mathbf{p}}_{0},\\
&\tilde{\mathbf{u}}_{k,p}=\left(\tilde T_{k,p}\right)^{-1}\mathbf{h}^H_k\tilde{\mathbf{p}}_{k}~\text{and}~\tilde v_{k,p}=1-\tilde{\mathbf{u}}^H_{k,p}\mathbf{h}^H_k\tilde{\mathbf{p}}_{k}.
\end{split}
\end{equation}
where $\tilde{\mathbf{p}}_{k}$ is the expansion point of beamfocusing vector $\mathbf{p}_{k}$ for $\forall k\in\tilde{\mathcal{K}}$. $\tilde T_{k,c}$ and $\tilde T_{k,p}$ are the received power at the expansion point, given by $\tilde T_{k,p}=\sum^{K}_{i=1}\left|\mathbf{h}^H_{k}\tilde{\mathbf{p}}_i\right|^2+\sigma^2_k$ and $\tilde T_{k,c}=\tilde T_{k,p}+\left|\mathbf{h}^H_{k}\tilde{\mathbf{p}}_0\right|^2$. The constructed surrogates $f_{k,p}\left(\mathbf{P}\right)$ form strict low bounds to the original logarithmic rate, satisfying $\log\left(1+\gamma_{k,\tau}\right)\geq f_{k,p}\left(\mathbf{P}\right)$ with equality holding at $\mathbf{P}=\tilde{\mathbf{P}}$. This proof shares a similar methodology with\cite{zhou2025sub}, to which readers are referred for detailed derivations. 

\subsubsection{Surrogate construction for eavesdropping rate} To guarantee the secrecy rate is concave, it is necessary to build upper-bounded convex surrogates for the eavesdropping rate. However, the construction method effective for the legitimate rates generates only lower-bounded surrogates for the eavesdropping rate. Similarly, classical techniques such as weighted minimum mean-squared error (WMMSE) and conventional quadratic transforms are no longer applicable, thereby necessitating an alternative solution framework. Herein, we first rewrite
$-R_{\text{e},c}$ as
\begin{align}
-R_{\text{e},c}&=-\log\left(1+\frac{\left|\mathbf{g}^H_\text{e}\mathbf{p}_0\right|^2}{\sum^{K}_{i=1}\left|\mathbf{g}^H_\text{e}\mathbf{p}_i\right|^2+\sigma^2_{\text{e}}}\right)\notag\\
&\overset{(a)}{=}\log\left(\frac{\sum^{K}_{i=1}\left|\mathbf{g}^H_\text{e}\mathbf{p}_i\right|^2+\sigma^2_{\text{e}}}{\sum^{K}_{i=0}\left|\mathbf{g}^H_\text{e}\mathbf{p}_i\right|^2+\sigma^2_{\text{e}}}\right)\notag\\
&\overset{(b)}{=}\log\left(\frac{\left|\mathbf{g}^H_\text{e}\mathbf{T}_0\right|^2}{T_{{\text{e}},c}}\right),
\end{align}
where 
\begin{equation}
\mathbf{T}_0=\left[\frac{\sigma^2_{\text{e}}}{N|\beta_{\text{e}}|^2}\mathbf{g}_{\text{e}},\mathbf{p}_{1},\dots,\mathbf{p}_K\right].
\end{equation}
Equality (a) invalidates the quadratic transform method\cite{shen2018fractional} unless an excessive number of auxiliary variables are introduced. In contrast, equality (b) enables the direct application of the quadratic transform to build a concave surrogate for $R_{\text{e},c}$, which is expressed as
\begin{align}
f_{{\text{e}},c}\left(\mathbf{x}_{\text{e},c},\mathbf{P}\right)= \log\left(2\text{Re}\left(\mathbf{x}^H_{\text{e},c}\mathbf{T}^H_0\mathbf{g}_\text{e}\right)-\mathbf{x}^H_{\text{e},c}T_{\text{e},c}\mathbf{x}_{\text{e},c}\right),\label{surrogate_common}
\end{align}
where $\mathbf{x}_{\text{e},c}\in\mathbb{C}^{N\times 1}$ is an auxiliary variable. Following the derivations in\cite{shen2018fractional}, we have
\begin{equation}
-R_{\text{e},c}= \max_{\mathbf{x}_{\text{e},c}}f_{{\text{e}},c}\left(\mathbf{x}_{\text{e},c},\mathbf{P}\right),
\label{equalivent}
\end{equation}
where the optimal solution to the right-hand of equation (\ref{equalivent}) is  
\begin{equation}
\mathbf{x}^*_{\text{e},c}=\frac{\mathbf{T}^H_0\mathbf{g}_\text{e}}{T_{\text{e},c}}.
\label{Optimal_common}
\end{equation}

Similarly, the eavesdropping rate $-R_{e,k}$ for $\forall k$ can be rewritten as $-R_{e,k}=\log\left(\frac{\left|\mathbf{g}^H_{\text{e}}\mathbf{T}_k\right|^2}{T_{{\text{e}},c}}\right)$, where 
\begin{equation}
\mathbf{T}_k=\left[\mathbf{p}_0,\dots,\mathbf{p}_{k-1},\frac{\sigma^2_{\text{e}}}{N|\beta_{\text{e}}|^2}\mathbf{g}_{\text{e}},\mathbf{p}_{k+1},\dots,\mathbf{p}_K\right].
\end{equation}
Its surrogate is 
\begin{align}
f_{{\text{e}},k}\left(\mathbf{x}_{\text{e},k},\mathbf{P}\right)=\log\left(2\text{Re}\left(\mathbf{x}^H_{\text{e},k}\mathbf{T}^H_k\mathbf{g}_\text{e}\right)-\mathbf{x}^H_{\text{e},k}T_{\text{e},c}\mathbf{x}_{\text{e},k}\right),\label{surrogate_private}
\end{align}
where the optimal $\mathbf{x}_{\text{e},k}$ is
\begin{equation}
\mathbf{x}^*_{\text{e},k}=\frac{\mathbf{T}^H_k\mathbf{g}_\text{e}}{T_{\text{e},c}}.
\label{Optimal_private}
\end{equation}

Based on the constructed surrogates (\ref{Overall_SCA}), (\ref{surrogate_common}), and (\ref{surrogate_private}), problem (\ref{linear_p4}) can be reformulated as
\begin{subequations}\label{linear_p5}
	\begin{align}
&\max_{\mathcal{Q}_1,\mathcal{Q}_2 } R^s -\frac{1}{\rho}||\mathbf{P}-\mathbf{FW}||^2_F,\label{ob_a5}\\
\text{s.t.}~&f_{k,c}\left(\mathbf{P}\right)+f_{{\text{e}},c}\left(\mathbf{x}_{\text{e},c},\mathbf{P}\right)\geq 0,~\forall k,\label{ob_b5}\\
    &R^s_{k,p}\leq f_{k,p}\left(\mathbf{P}\right)+f_{{\text{e}},k}\left(\mathbf{x}_{\text{e},k},\mathbf{P}\right),~\forall k,\label{ob_c5}\\
    &\sum_{k=1}^{K}R^c_{k,c} \leq f_{k,c}\left(\mathbf{P}\right)+f_{{\text{e}},c}\left(\mathbf{x}_{\text{e},c},\mathbf{P}\right),~\forall k,\label{ob_d5}\\
  &\mbox{ (\ref{ob_e}),~(\ref{ob_b3}),~(\ref{ob_c3})}.\label{ob_d5}
	\end{align}
\end{subequations}
where $\mathcal{Q}_2=\left\{\mathbf{x}_{\text{e},c},\mathbf{x}_{\text{e},k}\right\}$. The mutual dependence between $\mathcal{Q}_1$ and $\mathcal{Q}_2$ complicates joint optimization. However, we observe that all constraints reduce to convex sets for fixed $\mathcal{Q}_2$, whereas, for given $\mathcal{Q}_1$, the optimal $\mathcal{Q}_2$ can be derived in closed form via (\ref{Optimal_common}) and (\ref{Optimal_private}). Leveraging this structure, we adopt an alternating optimization framework that updates $\mathcal{Q}_1$ and $\mathcal{Q}_2$ alternately until convergence is achieved.  The complete procedure is outlined in Algorithm \ref{Alg.1}.
\begin{algorithm}[t]
	\caption{Iterative algorithm for solving (\ref{linear_p4})}
	\begin{algorithmic}[1]\label{Alg.1}
		\STATE Initialize a feasible $\mathbf{p}_k$ for $\forall k\in\mathcal{K}_1$. 
		\REPEAT
        \STATE Update $\tilde{\mathbf{p}}_{k}=\mathbf{p}_{k}$ for $\forall k\in\mathcal{K}_1$.
        \STATE Update $\mathcal{Q}_2$ based on equations (\ref{Optimal_common}) and (\ref{Optimal_private}).
		\STATE  Solving problem (\ref{linear_p5}) to obtain optimal $\mathbf{p}_{k}$.
		\UNTIL{the increment of the objective value of problem  (\ref{linear_p4}) falls below a predefined threshold.}	
		\STATE Return the optimized $\mathbf{P}$.
	\end{algorithmic}
\end{algorithm}

\subsection{Subproblem with respect to $\mathbf{W}$ } 
The digital beamfocuser $\mathbf{W}$ only appears in the last term of the objective function. Consequently, when $\mathbf{P}$ and $\mathbf{F}$ are fixed, problem (\ref{linear_p3}) reduces to the following unconstrained formulation:
\begin{equation}
\min_{\mathbf{W}}||\mathbf{P}-\mathbf{F}\mathbf{W}||^2_F,
\end{equation}
which admits a closed-form solution derived from the first-order optimality condition. The optimal digital beamfocuser is
\begin{align}\label{linear_p6}
\mathbf{W}^*=\left(\mathbf{F}^H\mathbf{F}\right)^{-1}\mathbf{F}^H\mathbf{P}.
\end{align}

\subsection{Subproblem with respect to $\mathbf{F}$ } 
\subsubsection{Fully-connected HAD architecture} With fixed $\mathbf{P}$ and $\mathbf{W}$,  problem (\ref{linear_p3}) under fully-connected HAD configuration can be reformulated to
\begin{subequations}\label{linear_p10}
	\begin{align}
&\min_{\mathbf{F} }\mathrm{Tr}\left(\mathbf{F}^H\mathbf{F}\mathbf{Y}\right)-2\mathrm{Re}\left(\mathrm{Tr}\left(\mathbf{F}^H\mathbf{Z}\right)\right),\label{ob_a10}\\
	\text{s.t.}~
 &|\mathbf{F}_{n,l}|=1,~\forall n,~\forall l,\label{ob_b10}
	\end{align}
\end{subequations}
where $\mathbf{Y}=\mathbf{W}\mathbf{W}^H$ and $\mathbf{Z}=\mathbf{P}\mathbf{W}^H$. It is observed that the unit-modulus constraint (\ref{ob_b10}) exhibits element-wise separability, thereby motivating the adoption of an element-wise optimization strategy. Accordingly, the subproblem for optimizing $\mathbf{F}_{n,l}$ is given by
\begin{subequations}\label{linear_p11}
	\begin{align}
&\min_{\mathbf{F}_{n,l}}\phi_{n,l}|\mathbf{F}_{n,l}|^2-2\text{Re}\left(\chi_{n,l}\mathbf{F}_{n,l}\right),\label{ob_a11}\\
\text{s.t.}~
 &|\mathbf{F}_{n,l}|=1,\label{ob_b11}
	\end{align}
\end{subequations}
where $\phi_{n,l}$ and $\chi_{n,l}$ denote real and complex constant coefficients, respectively, determined by all elements of $\mathbf{F}$ except $\mathbf{F}_{n,l}$. Enforcing the unit-modulus constraint $|\mathbf{F}_{n,l}|=1$, the optimal $\mathbf{F}_{n,l}$ is obtained as
\begin{equation}
\mathbf{F}^*_{n,l}=e^{-j\angle{\chi_{n,l}}}.
\label{Optimal_analog}
\end{equation} 
The exact solution remains unattainable at this stage, as the coefficient $\chi_{n,l}$ is unknown. Nevertheless, the objective functions (\ref{linear_p10}) and (\ref{linear_p11}) possess the same partial derivatives with respect to $\mathbf{F}_{n,l}$. Therefore, we have
\begin{equation}
\mathbf{X}_{n,l} -\mathbf{Z}_{n,l} = \phi_{n,l}\tilde{\mathbf{F}}_{n,l}-\chi_{n,l},
\end{equation} 
where $\mathbf{X} = \tilde{\mathbf{F}}\mathbf{Y}$, and $\tilde{\mathbf{F}}$ denotes the optimized solution of $\mathbf{F}$ from the previous iteration. Furthermore, by expanding $\tilde{\mathbf{F}}\mathbf{Y}$, we have $\phi_{n,l}\tilde{\mathbf{F}}_{n,l}=\tilde{\mathbf{F}}_{n,l}\mathbf{Y}_{l,l}$, leading to 
\begin{equation}
\chi_{n,l}=\mathbf{Z}_{n,l}-\mathbf{X}_{n,l} + \tilde{\mathbf{F}}_{n,l}\mathbf{Y}_{l,l}.
\label{Optimal_F}
\end{equation} 

\subsubsection{Sub-connected HAD architecture} Under the sub-connected configuration, analog beamfocuser $\mathbf{F}$ has a block-diagonal structure rather than a full matrix. Leveraging this property, the optimal analog beamfocusing matrix can be derived with closed-form expressions. In particular, the objective function can be expressed as
\begin{align}\label{New_obj_2}
&||\mathbf{P}-\mathbf{F}\mathbf{W}||^2_F=\sum^{L}_{l=1}||\bar{\mathbf{P}}_{l}-\mathbf{f}_{l}\bar{\mathbf{w}}_{l}||^2_F\notag\\
=&\bar\eta-\sum^{L}_{l=1}2\text{Re}\left(\bar{\mathbf{w}}_{l}\bar{\mathbf{P}}^H_l\mathbf{f}_{l}\right),
\end{align}
where
\begin{align}
&\bar{\mathbf{P}}_{l}=\mathbf{P}\big((l-1)M+1:lM,:\big),\notag\\
&\bar{\mathbf{w}}_{l} = \mathbf{W}(l,:),\notag\\
&\bar \eta =\sum^{L}_{l=1}\left(M\bar{\mathbf{w}}_{l}\bar{\mathbf{w}}^H_{l}+\text{Tr}\left(\bar{\mathbf{P}}_{l}\bar{\mathbf{P}}^H_{l}\right)\right).
\end{align}
Equation (\ref{New_obj_2}) reveals that $\min_{\mathbf{F}} ||\mathbf{P}-\mathbf{F}\mathbf{W}||^2_F$ can be decomposed into $L$ independent subproblems when $\mathbf{P}$ and $\mathbf{F}$ are fixed. For $\mathbf{f}_l$, its optimization subproblem is
\begin{subequations}\label{linear_e2}
	\begin{align}
&\max_{\mathbf{f}_{l}}\text{Re}\left(\bar{\mathbf{w}}_{l}\bar{\mathbf{P}}^H_l\mathbf{f}_{l}\right),\label{ob_ea2}\\
	\text{s.t.}~
  &|\mathbf{f}_{l,m}|=1,~\forall m.\label{ob_eb2}
	\end{align}
\end{subequations}
whose optimal solution is readily obtained as
\begin{equation}
\mathbf{f}^*_{l}=\left(e^{-j\angle\bar{\mathbf{w}}_{l}\bar{\mathbf{P}}^H_l}\right)^T.
\label{PS_beamfocusing_2}
\end{equation}

\subsection{Overall algorithm and properties analysis}
Based on the block-wise solutions, we summarize the penalty-based alternating optimization algorithm in Algorithm~\ref{Alg.2}. Its convergence and complexity are discussed below:
\begin{itemize}
\item  \emph{Convergence}: Starting from an arbitrary feasible initial point, Algorithm~\ref{Alg.1} and Algorithm~\ref{Alg.2} yield globally optimal solutions in lines $3\sim 5$ and lines $5\sim 6$, respectively. This reveals that Algorithm~\ref{Alg.2} identifies the previous feasible point at least in each iteration, thereby producing a non-decreasing sequence of objective values in lines $4\sim 6$ under a fixed penalty factor $\rho$. Specifically, we have
\begin{align}
&R^s\left(\mathcal{Q}^{(t)}_1,\mathbf{W}^{(t)},\mathbf{F}^{(t)}\right)\notag\\\geq& R^s\left(\mathcal{Q}^{(t)}_1,\mathbf{W}^{(t)},\mathbf{F}^{(t-1)}\right)\notag\\\geq& R^s\left(\mathcal{Q}^{(t)}_1,\mathbf{W}^{(t-1)},\mathbf{F}^{(t-1)}\right)\notag\\\geq& R^s\left(\mathcal{Q}^{(t-1)}_1,\mathbf{W}^{(t-1)},\mathbf{F}^{(t-1)}\right)
 \label{convergence}
\end{align}
Additionally, since the secrecy rate is upper-bounded, the inner loop of Algorithm~\ref{Alg.2} is guaranteed to converge within a finite number of iterations. Furthermore, the penalty-based method employed in the outer loop has been proven to converge to a stationary
point\cite{9120361}. Based on the above insights, we can deduce that Algorithm~\ref{Alg.2} converges.

\item \emph{Complexity}: In line 4, the computational load stems from solving problem (\ref{linear_p5}) to obtain optimal $\mathbf{P}$. With $V$ number of optimization variables, the complexity of the conventional interior point method is $\mathcal O\left(V^{3.5}\right)$. As such, the complexity of line 4 is $\mathcal O\left(\delta_1\left(N(K+1)+2K+1\right)^{3.5}\right)$, where $\delta_1$ denotes the iteration number until Algorithm~\ref{Alg.1} converges. For two matrices $\mathbf{W}_1\in\mathbb C^{ A_1\times A_2}$ and $\mathbf{W}_2\in\mathbb C^{ A_2\times A_3}$, the complexity of $\mathbf{W}_1\mathbf{W}_2$ is $\mathcal O\left(A_1A_2A_3\right)$. Therefore, the complexity of lines 5 is in order of $\mathcal O\left(NL\max(L,K+1)\right)$ for the fully-connected architecture, and $\mathcal O\left(NL(K+1)\right)$ for the sub-connected architecture. The complexity of lines 6 is in order of $\mathcal O\left(NL(K+1)^2\right)$.
\end{itemize}

\begin{algorithm}[t]
	\caption{Penalty-based alternating optimization algorithm for solving (\ref{linear_p3})}
	\begin{algorithmic}[1]\label{Alg.2}
		\STATE Initialize $\mathbf{F}$ and $\mathbf{W}$.
        \REPEAT
		\REPEAT
		\STATE  Update $\mathbf{P}$ by invoking Algorithm~\ref{Alg.1}.
        \STATE Update $\mathbf{W}$ according to (\ref{linear_p6}).
        \STATE Update $\mathbf{F}$ according to using (\ref{Optimal_F}) (fully-connected) or according to (\ref{PS_beamfocusing_2}) (sub-connected).
		\UNTIL { the increment of the objective value of problem (\ref{linear_p3}) falls below a predefined threshold. }
        \STATE Update penalty factor $\rho=\alpha  \rho$.
        \UNTIL{the penalty term falls below a predefined threshold. }
		\STATE Return the optimized max-min secrecy rate $R^s$.
	\end{algorithmic}
\end{algorithm}

\section{Simulation results}\label{Section IV}
This section presents numerical results to evaluate the secrecy performance of the proposed transmit scheme. Unless otherwise specified, the simulation parameters are configured as follows. The BS is equipped with $N=128$ antennas and $L=8$ RF chains with half-wavelength spacing, operating at a carrier frequency of $f_c=30$~GHz. $K=4$ legitimate users and one eavesdropper are randomly generated within a distance range of $[10,20]$ meters and an angular range of $[0,\tfrac{\pi}{2}]$. The maximum transmit power and noise power are $P_{\text{th}}=20$~dBm and $\sigma^2_{\text{e}}=\sigma^2_k=-84$~dBm.  The penalty factor is initialized as  $\rho=10^2$ with a reduction factor of $\alpha=0.5$. These settings are primarily taken from\cite{10579914,zhou2025sub}.

Over $100$ independent spherical wave channel realizations,  we simulate our proposed RSMA-enhanced secure transmit schemes with fully-connected and sub-connected architectures, labeled as {\bf{RSMA-FC}} and {\bf{RSMA-SC}}, respectively.  For a  comprehensive evaluation, we benchmark them against the following four baselines:
\begin{itemize}
\item {\bf{RSMA-FD}}: The BS employs fully digital beamfocusing, where the number of RF  chains $L$ equals the number of transmit antennas $N$. It provides the secrecy rate upper bound for our proposed HAD architectures. 
\item {\bf{RSMA-Comm}}: This baseline ignores eavesdropping by assuming no eavesdroppers, thereby reducing the optimization objective to maximizing the minimum transmit rate. The obtained results reveals the impact of eavesdropping on the users' transmit rates, while the observed performance gap quantifies the beamfocusing capability.
\item {\bf{SDMA-FC}}:  This baseline relies solely on near-field beamfocusing under the fully-connected HAD architecture to counter eavesdropping. Specifically, each message is encoded into a dedicated private stream (i.e., $\mathbf{w}_0=\mathbf{0}$), and each user directly decodes its desired stream by treating interference as noise.

\item {\bf{RSMA-FC-far}}: This benchmark adopts the plane wave-based far-field channel model and fully-connected HAD architecture, where the array response vector for user $i$ is given by
\begin{align}
\mathbf{a}_{\text{far}}\left(\theta_i\right)= \left[e^{j\frac{2\pi}{\lambda}y_1\sin\theta_i},\dots,e^{j\frac{2\pi}{\lambda}y_N\sin\theta_i}\right]^T
\label{Far-Channel}
\end{align}
with $\forall i\in\left\{1,\dots,K,\text{e}\right\}$. Except for the channel model, all other parameters remain unchanged to ensure fair comparison.
\end{itemize}

\begin{figure}[tbp]
	\centering
	\includegraphics[scale=0.6]{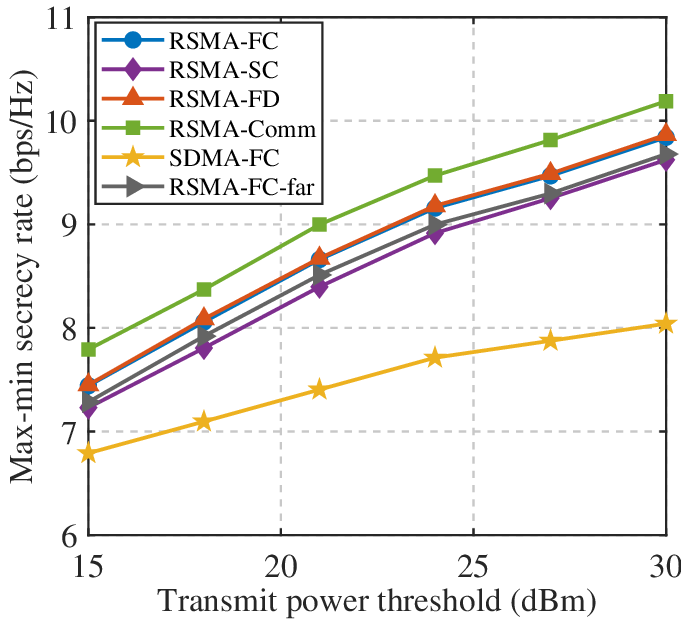}
	\caption{Max-min secrecy rate versus the transmit power threshold.}
	\label{power}
\end{figure}
Fig.~\ref{power} illustrates the max-min secrecy rate versus the transmit power threshold, highlighting three key observations over four benchmark schemes.
\begin{enumerate}
    \item\emph{Effective eavesdropping suppression}: The gap between the max-min communication rate and the max-min secrecy rate is approximately $0.3$~bps/Hz, indicating that the maximum eavesdropping rate is effectively limited. This validates the dual functionality of the RSMA common stream in delivering intended messages while impairing eavesdroppers.
    \item\emph{Imperfect beamfocusing capability}: The proposed secure framework surpasses far-field beamforming and beamfocusing-only transmit scheme, with the performance gap widening as the transmit power increases. This demonstrates that near-field spherical waves effectively mitigate energy leakage, although imperfect beamfocusing prevents complete elimination.
    \item\emph{Higher hardware efficiency}: Using only $8$ RF chains, our proposed HAD antenna architectures, especially the fully-connected configuration, achieve performance consistently close to fully digital beamfocusing across all transmit power levels, thereby demonstrating their hardware efficiency.
\end{enumerate}

\begin{figure}[tbp]
	\centering
	\includegraphics[scale=0.6]{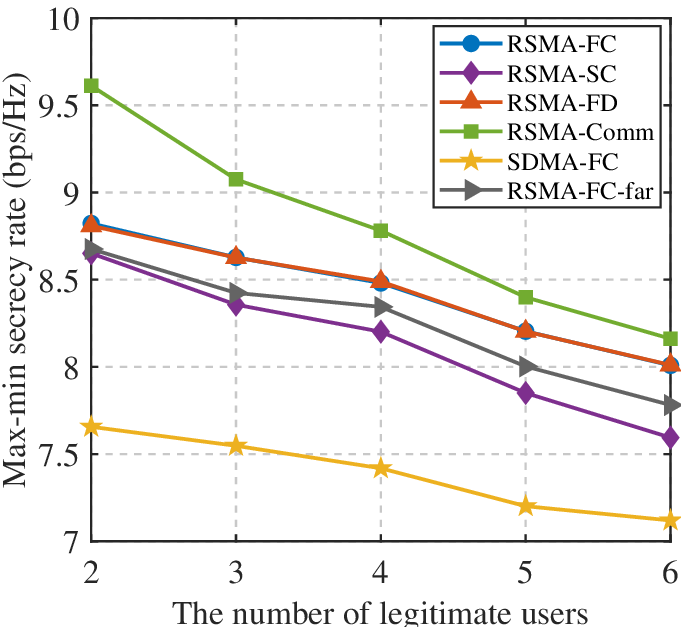}
	\caption{Max-min secrecy rate versus the number of legitimate users.}
	\label{user}
\end{figure}
Fig.~\ref{user} presents the max-min secrecy rate versus the number of legitimate users. As expected, the max-min secrecy rate decreases for all transmit schemes as more legitimate users are scheduled. Moreover, the performance difference between fully digital beamfocusing and sub-connected HAD beamfocusing/far-field beamforming becomes more pronounced. This arises because sub-connected architectures reduce beamfocusing precision, while far-field beamforming lacks directional discrimination. This results in stronger multi-user interference and thus reduces communication rates. In addition, the difference between the communication rate and the secrecy rate reaches $0.8$~bps/Hz when $K=2$. Two main factors contribute to this phenomenon. First, the common rate depends on the user with the poorest channel quality and is shared by all legitimate users. Second, the eavesdropper experiences lower interference when fewer legitimate users are scheduled, enhancing its decoding capability. 

\begin{figure}[tbp]
	\centering
	\includegraphics[scale=0.6]{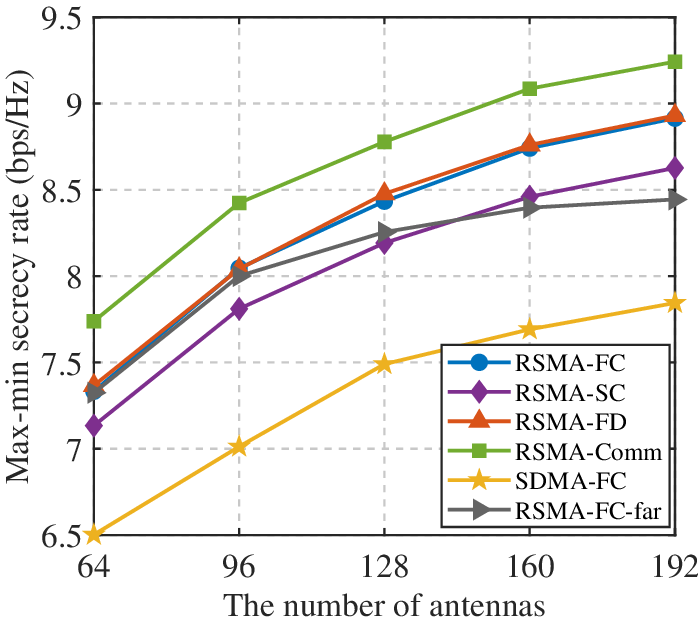}
	\caption{Max-min secrecy rate versus the number of antennas.}
	\label{antenna}
\end{figure}
Fig.~\ref{antenna} depicts the max-min secrecy rate as a function of the number of antennas. As anticipated, all schemes benefit from larger antenna arrays, which provide higher spatial degrees of freedom and improved beamfocusing. Compared with the beamfocusing-only scheme, our proposed secure transmit scheme achieves an approximate $1$~bps/Hz gain, highlighting the RSMA common stream’s effectiveness in interference mitigation and eavesdropping resistance. Besides, the performance gap between near-field beamfocusing and far-field beamforming gradually widens as antenna array size grows, reaching $0.5$~bps/Hz at $N=192$. Interestingly, the fully connected far-field beamfocusing scheme becomes inferior to the sub-connected near-field beamfocusing scheme when $N\geq 128$. This underscores the enhanced signal strength and energy leakage suppression capabilities provided by near-field beamfocusing. 

\begin{figure}[tbp]
	\centering
	\includegraphics[scale=0.6]{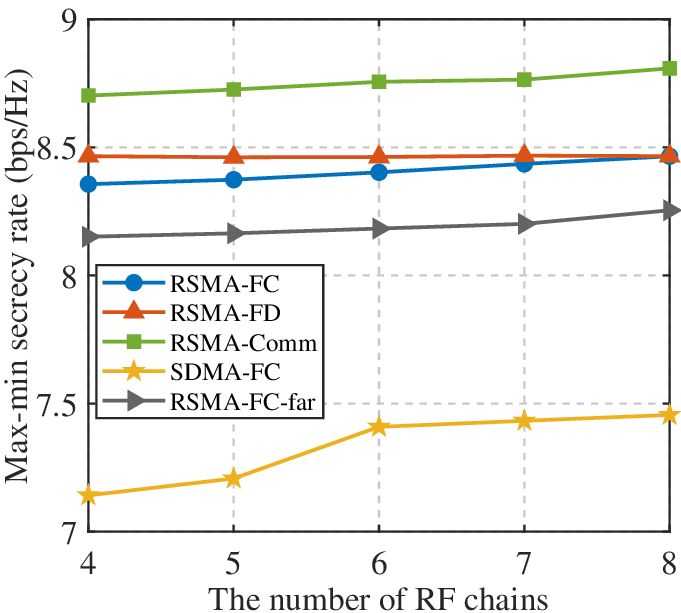}
	\caption{Max-min secrecy rate versus the number of RF chains.}
	\label{RF}
\end{figure}
Fig.~\ref{RF} investigates the impact of the number of RF chains on the max-min secrecy rate. The performance of fully digital beamfocusing remains horizon, as it is independent of the number of RF chains, serving as an upper-bound reference. The performance gap between fully connected HAD and fully digital beamfocusing gradually diminishes. Even when $K=L$, the performance loss is below $0.1$~bps/Hz, while the required number of RF chains is reduced by a factor of $32$. Furthermore, when $L=2K$, the gap becomes almost negligible. However, under the same number of RF chains, the proposed secure transmit scheme consistently outperforms both far-field beamforming and beamfocusing-only schemes, demonstrating its practical advantages for secure NFC. 

\begin{figure}[tbp]
	\centering
	\includegraphics[scale=0.6]{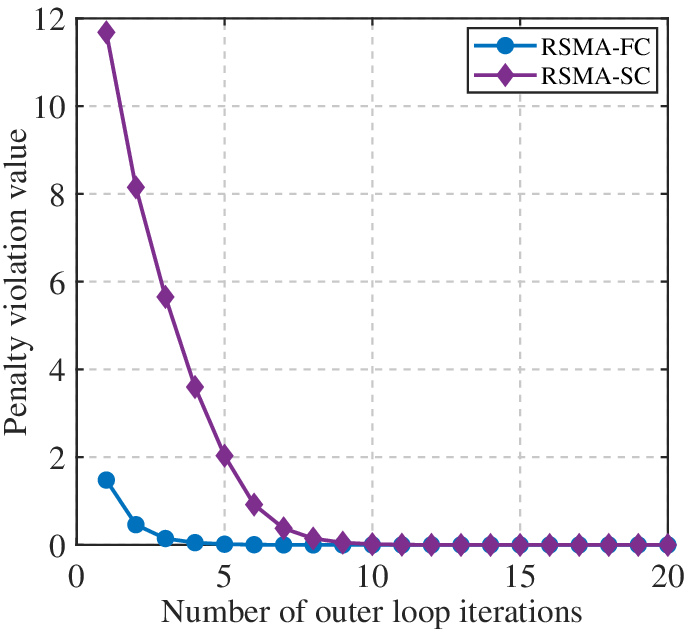}
	\caption{Convergence behavior of the proposed algorithms.}
	\label{convergence}
\end{figure}
Fig.~\ref{convergence} illustrates the variation of the penalty violation term $||\mathbf{P}-\mathbf{FW}||^2_F$ across outer loop iterations. The result validates that the penalty violation converges to zero within a limited number of iterations: approximately $5$ for the fully-connected and $10$ for the sub-connected HAD architecture, respectively. Notably, even with a penalty coefficient as low as $1/100$, the penalty violation value remains below $2$. This finding, in turn, serves as additional validation that the fully-connected HAD architecture with $8$ RF chains can achieve performance nearly comparable to that of the fully-digital counterpart.

\section{Conclusion}\label{Section V}
This paper proposes an RSMA-enhanced secure transmit scheme, where the common stream simultaneously conveys the intended message to legitimate users and acts as artificial noise to impair eavesdroppers. The analog beamfocuser, digital beamfocuser, and common secrecy rate allocation are jointly optimized to maximize the minimum secrecy rate. To addressing the resulting non-convex problem, we develop a penalty-based alternating optimization algorithm, in which the optimal analog and digital beamfocusers are derived in closed form. Numerical results demonstrate that beamfocusing alone is insufficient to fully suppress energy leakage.
Moreover, the proposed secure scheme achieves significantly higher secrecy performance than far-field beamforming, while approaching the performance of fully digital beamfocusing with substantially fewer RF chains.

	\ifCLASSOPTIONcaptionsoff
	\newpage
	\fi
	
	\bibliographystyle{IEEEtran}
	\bibliography{references}

@ARTICLE{10478949,
  author={Jiang, Xiao and Li, Peng and Shang, Yuling and Zou, Yulong and Li, Bin and Yan, Peishun},
  journal={{IEEE} Trans. Commun.}, 
  title={Improving Physical Layer Security for Distributed Antenna Systems With a Friendly Jammer}, 
  year={2024},
  month={Aug.},
  volume={72},
  number={8},
  pages={4756-4773}}

@ARTICLE{10311400,
  author={Xiang, Luping and Zeng, Yao and Hu, Jie and Yang, Kun and Hanzo, Lajos},
  journal={{IEEE} Trans. Commun.}, 
  title={Multi-Domain Polarization for Enhancing the Physical Layer Security of {MIMO} Systems}, 
  year={2024},
  month={Mar.},
  volume={72},
  number={3},
  pages={1502-1515}}

@ARTICLE{10285055,
  author={Zhou, Jiasi and Hou, Wenjun and Mao, Yijie and Tellambura, Chintha},
  journal={{IEEE} Commun. Lett.}, 
  title={Artificial Noise Assisted Secure Transmission for Uplink {MIMO} Rate Splitting Healthcare Systems}, 
  year={2023},
  month={Dec.},
  volume={27},
  number={12},
  pages={3176-3180}}

@ARTICLE{9792580,
  author={Zheng, Tong-Xing and Wen, Yating and Liu, Hao-Wen and Ju, Ying and Wang, Hui-Ming and Wong, Kai-Kit and Yuan, Jinhong},
  journal={{IEEE} Trans. Wireless Commun.}, 
  title={Physical-Layer Security of Uplink mm{W}ave Transmissions in Cellular {V2X} Networks}, 
  year={2022},
  month={Nov.},
  volume={21},
  number={11},
  pages={9818-9833}}

@ARTICLE{10558818,
  author={An, Jiancheng and Yuen, Chau and Dai, Linglong and Di Renzo, Marco and Debbah, Mérouane and Hanzo, Lajos},
  journal={{IEEE} Wirel. Commun.}, 
  title={Near-Field Communications: Research Advances, Potential, and Challenges}, 
  year={2024},
month={Jun.},
  volume={31},
  number={3},
  pages={100-107}}

@ARTICLE{10663521,
  author={Cong, Jiayi and You, Changsheng and Li, Jiapeng and Chen, Li and Zheng, Beixiong and Liu, Yuanwei and Wu, Wen and Gong, Yi and Jin, Shi and Zhang, Rui},
  journal={{IEEE} Wirel. Commun.}, 
  title={Near-Field Integrated Sensing and Communication: Opportunities and Challenges}, 
  year={2024},
  month={Dec.},
  volume={31},
  number={6},
  pages={162-169}}

@ARTICLE{10436390,
  author={Zhang, Zheng and Liu, Yuanwei and Wang, Zhaolin and Mu, Xidong and Chen, Jian},
  journal={{IEEE} Trans. Veh. Technol.}, 
  title={Physical Layer Security in Near-Field Communications}, 
  year={2024},
  month={Jul.},
  volume={73},
  number={7},
  pages={10761-10766}}

@ARTICLE{10123941,
  author={Wu, Zidong and Dai, Linglong},
  journal={{IEEE} J. Select. Areas Commun.}, 
  title={Multiple Access for Near-Field Communications: {SDMA} or {LDMA}?}, 
  year={2023},
  month={Jun.},
  volume={41},
  number={6},
  pages={1918-1935}}

@article{zhou2025sub,
  title={Sub-Connected Hybrid Beamfocusing Design for {RSMA}-Enabled Near-Field Communications with Imperfect {CSI} and {SIC}},
  author={Zhou, Jiasi and Chen, Ruirui and Sun, Yanjing and Tellambura, Chintha},
  journal={arXiv preprint arXiv:2507.11854},
  year={2025}}

@article{zhao2025near,
  title={Near-Field Integrated Sensing and Communications for Secure {UAV} Networks},
  author={Zhao, Jingjing and Xue, Songtao and Cai, Kaiquan and Mu, Xidong and Liu, Yuanwei and Zhu, Yanbo},
  journal={arXiv preprint arXiv:2502.01003},
  year={2025}}

@ARTICLE{10480457,
  author={Chen, Jiangong and Xiao, Yue and Liu, Kanglai and Zhong, Yuan and Lei, Xia and Xiao, Ming},
  journal={{IEEE} Trans. Veh. Technol.}, 
  title={Physical Layer Security for Near-Field Communications via Directional Modulation}, 
  year={2024},
  month={Aug.},
  volume={73},
  number={8},
  pages={12242-12246}}

@ARTICLE{10415047,
  author={Zhou, Jiasi and Hou, Wenjun and Mao, Yijie and Tellambura, Chintha},
  journal={{IEEE} Commun. Lett.}, 
  title={Securing Medical Sensor Data: A Novel Uplink Scheme With Rate Splitting and Active Intelligent Reflecting Surface}, 
  year={2024},
month={Mar.},
  volume={28},
  number={3},
  pages={493-497}}

@ARTICLE{10273395,
  author={Park, Jeonghun and Lee, Byungju and Choi, Jinseok and Lee, Hoon and Lee, Namyoon and Park, Seok-Hwan and Lee, Kyoung-Jae and Choi, Junil and Chae, Sung Ho and Jeon, Sang-Woon and Kwak, Kyung Sup and Clerckx, Bruno and Shin, Wonjae},
  journal={{IEEE} Netw.}, 
  title={Rate-Splitting Multiple Access for {6G} Networks: Ten Promising Scenarios and Applications}, 
  year={2024},
month={May},
  volume={38},
  number={3},
  pages={128-136}}

@article{mao2018rate,
  title={Rate-splitting multiple access for downlink communication systems: Bridging, generalizing, and outperforming {SDMA} and {NOMA}},
  author={Mao, Yijie and Clerckx, Bruno and Li, Victor OK},
  journal={EURASIP J. Wireless Commun. Netw.},
  volume={2018},
  number={1},
  pages={1--54},
  month={May},
  year={2018}}

@ARTICLE{10945425,
  author={Liu, Ziwei and Chen, Wen and Wu, Qingqing and Li, Zhendong and Zhu, Xusheng and Wu, Qiong and Cheng, Nan},
  journal={{IEEE} Trans. Commun.}, 
  title={Enhancing Robustness and Security in {ISAC} Network Design: Leveraging Transmissive Reconfigurable Intelligent Surface with {RSMA}}, 
  year={2025},
  volume={},
  number={},
  pages={1-1},
  note={doi={10.1109/TCOMM.2025.3555894}}}

@article{zhang2025fluid,
  title={Fluid Antenna-Aided Robust Secure Transmission for {RSMA}-{ISAC} Systems},
  author={Zhang, Cixiao and Xu, Yin and Peng, Size and Guo, Xinghao and Ou, Xiaowu and Hong, Hanjiang and He, Dazhi and Zhang, Wenjun},
  journal={arXiv preprint arXiv:2503.05515},
  year={2025}
}

@ARTICLE{10971913,
  author={Zhao, Boqun and Ouyang, Chongjun and Zhang, Xingqi and Liu, Yuanwei},
  journal={{IEEE} Trans. Wireless Commun.}, 
  title={Performance Analysis of Physical Layer Security: From Far-Field to Near-Field}, 
  year={2025},
  volume={},
  number={},
  pages={1-1},
  note={doi={10.1109/TWC.2025.3560568}}}

@article{liu2025physical,
  title={Physical-Layer Security in Mixed Near-Field and Far-Field Communication Systems},
  author={Liu, Tianyu and You, Changsheng and Zhou, Cong and Zhang, Yunpu and Gong, Shiqi and Liu, Heng and Zhang, Guangchi},
  journal={arXiv preprint arXiv:2504.19555},
  year={2025}}

@article{zhang2024near,
  title={Near-field wideband secure communications: An analog beamfocusing approach},
  author={Zhang, Yuchen and Zhang, Haiyang and Xiao, Sa and Tang, Wanbin and Eldar, Yonina C},
  journal={{IEEE} Trans. Signal Process.},
  volume={72},
  pages={2173--2187},
  month={Apr.},
  year={2024},
  publisher={IEEE}}

@ARTICLE{10902048,
  author={Lei, Jiayi and Mu, Xidong and Zhang, Tiankui and Liu, Yuanwei},
  journal={{IEEE} Trans. Veh. Technol.}, 
  title={{RIS} Assisted Near-Field {NOMA} Communications: A Security-Fairness Trade-Off}, 
  year={2025},
  month={Jul.},
  volume={74},
  number={7},
  pages={11656-11661}}

@ARTICLE{11018844,
  author={Xu, Yiming and Zheng, Mingxuan and Xu, Dongfang and Song, Shenghui and Da Costa, Daniel Benevides},
  journal={{IEEE} Trans. Wireless Commun.}, 
  title={Sensing-aided Near-Field Secure Communications with Mobile Eavesdroppers}, 
  year={2025},
  volume={},
  number={},
  pages={1-1},
  note={doi={10.1109/TWC.2025.3572688}}}

@ARTICLE{10646391,
  author={Tang, Kun and Wang, Zhengwu and Zheng, Beixiong and Feng, Wenjie and Che, Wenquan and Xue, Quan},
  journal={{IEEE} Wireless Commun. Lett.}, 
  title={{RSMA}-Enhanced Secure Transmission in {IRS}-Assisted Networks Against Internal and External Eavesdroppers}, 
  year={2024},
  month={Dec.},
  volume={13},
  number={12},
  pages={3310-3314}}

@ARTICLE{10542650,
  author={Xia, Huiyun and Mao, Yijie and Zhou, Xiaokang and Clerckx, Bruno and Han, Shuai and Li, Cheng},
  journal={{IEEE} Trans. Wireless Commun.}, 
  title={Weighted Sum-Rate Maximization of Rate-Splitting Multiple Access With Confidential Messages}, 
  year={2024},
  month={Oct.},
  volume={23},
  number={10},
  pages={13738-13751}}

@ARTICLE{10720715,
  author={Zhang, Yao and Zhao, Haitao and Xia, Wenchao and Zhu, Yongxu and Ngo, Hien Quoc and Tan, Bo},
  journal={{IEEE} Trans. Wireless Commun.}, 
  title={Enhancing Secrecy in Hardware-Impaired Cell-Free Massive {MIMO} by {RSMA}}, 
  year={2024},
  month={Dec.},
  volume={23},
  number={12},
  pages={18788-18805}}

@article{salem2024robust,
  title={Robust Secure {ISAC}: How {RSMA} and Active {RIS} Manage Eavesdropper's Spatial Uncertainty},
  author={Salem, A Abdelaziz and Abdallah, Saeed and Saad, Mohamed and Alnajjar, Khawla and Albreem, Mahmoud A},
  journal={arXiv preprint arXiv:2407.15113},
  year={2024}
}

@ARTICLE{10817512,
  author={Wang, Dawei and Li, Jiawei and Lv, Qinyi and He, Yixin and Li, Li and Hua, Qiaozhi and Alfarraj, Osama and Zhang, Jiankang},
  journal={{IEEE} Internet Things J.}, 
  title={Integrating Reconfigurable Intelligent Surface and {AAV} for Enhanced Secure Transmissions in {IoT}-Enabled {RSMA} Networks}, 
  year={2025},
  month={Apr.},
  volume={12},
  number={8},
  pages={9405-9419}}

@ARTICLE{10414053,
  author={Zheng, Guangyuan and Wen, Miaowen and Wen, Jinming and Shan, Chun},
  journal={{IEEE} Wireless Commun. Lett.}, 
  title={Joint Hybrid Precoding and Rate Allocation for {RSMA} in Near-Field and Far-Field Massive {MIMO} Communications}, 
  year={2024},
  month={Apr.},
  volume={13},
  number={4},
  pages={1034-1038}}

@ARTICLE{10798456,
  author={Zhou, Jiasi and Zhou, Cong and Mao, Yijie and Tellambura, Chintha},
  journal={{IEEE} Wireless Commun. Lett.}, 
  title={Joint Beam Scheduling and Resource Allocation for Flexible {RSMA}-Aided Near-Field Communications}, 
  year={2025},
moth={Feb.},
  volume={14},
  number={2},
  pages={554-558}}

@article{zhou2024hybrid,
  title={Hybrid Beamforming Design for {RSMA}-enabled Near-Field Integrated Sensing and Communications},
  author={Zhou, Jiasi and Zhou, Cong and Tellambura, Chintha and Li, Geoffrey Ye},
  journal={arXiv preprint arXiv:2412.17062},
  year={2024}
}

@ARTICLE{10906379,
  author={Zhou, Jiasi and Zhou, Cong and Zeng, Cheng and Tellambura, Chintha},
  journal={{IEEE} Trans. Veh. Technol.}, 
  title={Flexible Rate-Splitting Multiple Access for Near-Field Integrated Sensing and Communications}, 
  year={2025},
  month={Jul.},
  volume={74},
  number={7},
  pages={11524-11528},}

@article{zhou2025crb,
  title={{CRB}-Rate Tradeoff in {RSMA}-enabled Near-Field Integrated Multi-Target Sensing and Multi-User Communications},
  author={Zhou, Jiasi and Zhou, Cong and Sun, Yanjing and Tellambura, Chintha},
  journal={arXiv preprint arXiv:2502.11516},
  year={2025}}

@ARTICLE{10579914,
  author={Li, Haochen and Wang, Zhaolin and Mu, Xidong and Zhiwen, Pan and Liu, Yuanwei},
  journal={{IEEE} J. Select. Areas Commun.}, 
  title={Near-Field Integrated Sensing, Positioning, and Communication: A Downlink and Uplink Framework}, 
  year={2024},
  month={Sept.},
  volume={42},
  number={9},
  pages={2196-2212}}

@ARTICLE{8709756,
  author={Taghizadeh, Omid and Neuhaus, Peter and Mathar, Rudolf and Fettweis, Gerhard},
  journal={{IEEE} Trans. Commun.}, 
  title={Secrecy Energy Efficiency of {MIMOME} Wiretap Channels With Full-Duplex Jamming}, 
  year={2019},
  month={Aug.},
  volume={67},
  number={8},
  pages={5588-5603}}

@article{lange2000optimization,
  title={Optimization transfer using surrogate objective functions},
  author={Lange, Kenneth and Hunter, David R and Yang, Ilsoon},
  journal={J. Comput. Graph. Statist.},
  volume={9},
  number={1},
  pages={1--20},
  month={Mar.},
  year={2000},
  publisher={Taylor \& Francis}
}

@ARTICLE{8976409,
  author={Li, Yang and Xia, Minghua and Wu, Yik-Chung},
  journal={{IEEE} Trans. Wireless Commun.}, 
  title={Caching at Base Stations With Multi-Cluster Multicast Wireless Backhaul via Accelerated First-Order Algorithms}, 
  year={2020},
  month={May},
  volume={19},
  number={5},
  pages={2920-2933}}

@article{shen2018fractional,
  author={Shen, Kaiming and Yu, Wei},
  journal={{IEEE} Trans. Signal Process.}, 
  title={Fractional Programming for Communication Systems-Part {I}: Power Control and Beamforming}, 
  year={2018},
  volume={66},
  number={10},
  pages={2616-2630},
  month={May},
  doi={10.1109/TSP.2018.2812733}}

@ARTICLE{9120361,
  author={Shi, Qingjiang and Hong, Mingyi},
  journal={{IEEE} Trans. Signal Process.}, 
  title={Penalty Dual Decomposition Method for Nonsmooth Nonconvex Optimization—Part {I}: Algorithms and Convergence Analysis}, 
  year={2020},
month={Jun.},
  volume={68},
  number={},
  pages={4108-4122}}

@ARTICLE{11071287,
  author={Zhang, Shengyu and Wang, Feng and Mao, Yijie and Jin, A-Long and Quek, Tony Q.S.},
  journal={{IEEE} Trans. Commun.}, 
  title={Rate-Splitting Multiple Access for Near-Field Communications with Imperfect {CSIT} and {SIC}}, 
  year={2025},
  volume={},
  number={},
  pages={1-1},
  note={doi={10.1109/TCOMM.2025.3585513}}}

@ARTICLE{7555358,
  author={Joudeh, Hamdi and Clerckx, Bruno},
  journal={{IEEE} Trans. Commun.}, 
  title={Sum-Rate Maximization for Linearly Precoded Downlink Multiuser {MISO} Systems With Partial {CSIT}: A Rate-Splitting Approach}, 
  year={2016},
  month={Nov.},
  volume={64},
  number={11},
  pages={4847-4861}}
	
\end{document}